\documentclass[aps,pre,twocolumn]{revtex4-1}
\usepackage{graphicx}
\usepackage{amsmath}
\usepackage{amssymb}
\usepackage{braket}
\usepackage{hyperref}
\usepackage{subfigure}
\usepackage[usenames,dvipsnames]{color}

\usepackage{graphicx}
\usepackage{subfigure}
\usepackage[mathscr]{euscript}

\newcommand{\ct}{\cite}
\newcommand{\la}{\langle}
\newcommand{\bi}{\bibitem}
\newcommand{\be}{\begin{equation}}
\newcommand{\ee}{\end{equation}}
\newcommand{\ba}{\begin{eqnarray}}
\newcommand{\ea}{\end{eqnarray}}

\newcommand{\non}{\nonumber}
\newcommand{\Tr}{{\rm Tr}}

\newcommand{\Pq}{P_{\mathrm{q}}}
\newcommand{\Pqm}{P_{\mathrm{-q}}}

\newcommand{\ra}{\rangle}

\begin{document}

\title{Exploring the role of asymmetric-pulse modulation in quantum thermal machines and quantum thermometry}
\author{Saikat Mondal}
\author{Sourav Bhattacharjee}
\email{bsourav@iitk.ac.in} 
\author{Amit Dutta}

\affiliation{Department of Physics, Indian Institute of Technology Kanpur, Kanpur 208016, India}

\date{\today}

\begin{abstract}
We explore the consequences of periodically modulating a quantum two-level system (TLS) with an asymmetric pulse, when the system is in contact with thermal baths. By adopting the Floquet-Lindblad formalism for our analysis, we find that the unequal `up' and `down' time duration of the pulse has two main ramifications. Firstly, the energy gap of the multiple side-bands or photon sectors created as a result of the  periodic modulation are renormalized by a term which is dependent on both the modulation strength as well as the fraction of up (or down) time duration. Secondly, the weights of the different side-bands are no longer symmetrically distributed about the central band or zero photon sector. We illustrate the advantages of these findings in the context of applications in quantum thermal machines and thermometry. For a thermal machine constructed by coupling the TLS to two thermal baths, we demonstrate that the asymmetric pulse provides an extra degree of control over the mode of operation of the thermal machine. Further, by appropriately tuning the weight of the sub-bands, we also show that an asymmetric pulse may provide superior optimality in a recently proposed protocol for quantum thermometry, where dynamical control has been shown to enhance the precision of measurement.
	
\end{abstract}
\maketitle

\section{Introduction}\label{sec_intro}	
Since the early 2000s there has been a spurt in research on the modeling and dynamics of open quantum systems \cite{breuer02, rivas12, rotter15, breuer16}. These systems, in general, exhibit dissipative phenomena and therefore provide natural test-beds for investigating the quantum origin of dynamical processes such as quantum heat exchanges and quantum heat transport. A crucial ramification of this development has been the revitalization of the age-old quest for integrating quantum mechanics and thermodynamics into a single framework, thus leading to the emergence of the field of quantum thermodynamics \cite{alickiintro, kosloff13, anders16, esposito17, campaioli18, campbell19, kosloffrev19}. A major chunk of research in this field has focused on the building and analysing toy models of `quantum thermal machines', such as quantum equivalents of heat engines and refrigerators \cite{levy14, kurizki15, rezek17, mitchison19, bhattacharjee_review}; the motivation being that understanding the classical version of these machines had led to the foundation of classical thermodynamics.

As in classical thermal machines, the design and working of the quantum thermal machines are also based on cyclic processes, i.e. the system and the environment should return to their initial configuration. In this regard, {the Floquet-Lindblad formalism} of periodically driven open quantum systems has been formulated \cite{zanardi06, alickitut,szczyg13}to deal with quantum systems which, apart from being coupled with external environment, are described by a periodic time-dependent Hamiltonian. 
Importantly, a direct application of this framework has been shown in the context of designing continuous quantum thermal machines \cite{kosloff13, kurizki15, kurizki13,alicki14}. In the simplest of realizations, 
the working substance of these machines consist of a quantum system (with discrete energy levels) which is perpetually coupled to one or more baths or heat reservoirs{; this is in} contrast to reciprocating thermal machines, such as those based on Carnot or Otto cycles \cite{rezek17,kosloff92, bender00, nori07, tang09, uzdin15}, {where the baths couple to the system intermittently}. In addition, the Hamiltonian of the system is periodically modulated with the consequent dynamics resulting in exchange of quantum `heat' and `work'. A plethora of works \cite{kurizki15, klimovsky15, niedenzu15, weitz15, seifert17, mukherjee20} have explored the performance of these quantum machines in terms of work output and  efficiency, which have remarkably been found to be consistent with thermodynamic principles. On the other hand, {the Floquet-Linblad formalism has also revealed that periodic modulation of a `quantum probe' coupled to a thermal bath can significantly} enhance the precision \cite{mukherjee19} in low-temperature quantum thermometry \cite{correa15, pasquale16, brunner17, wang20, ancheyta20}  by increasing the relevant quantum Fisher information (QFI) \cite{caves94, paris09}, which in turn lowers the theoretical minimum relative error bound set by the quantum Cramer-Rao bound. 

However, all of the works mentioned above are based on a symmetric form of the periodic modulations. In other words, the form of the modulation on either side of the half-cycle are mirror copies of each other, for example -- sinusoidal or square-pulse modulations. In this work, we explore the consequences of using an asymmetric-pulse modulation (APM) in the two different but related applications discussed above, i.e. in designing quantum thermal machines and enhancement of precision in low-temperature quantum thermometry. The APM we use is of the form of a rectangular pulse whose `up' and `down' time within a single time-period are not necessarily equal (see Fig.~\ref{fig_asym_pulse}).

The motivation behind using an asymmetric modulation  is as follows. The Floquet-Lindblad approach \cite{zanardi06, alickitut} to the dynamics of periodically driven quantum system shows that the action of a Markovian thermal bath acting on a periodically driven system can be considered equivalent to infinite number of `sub-baths' acting simultaneously on the system. The multiple sub-baths are artifacts of the infinite number of `Floquet side-bands' or `photon sectors' created as a consequence of the periodic modulation, each of which independently exchange energies with the `physical baths'. It is worthwhile to note that each of the sub-baths, when acting independently on the system, steers the system towards different Gibbs states as determined by the energy-gap of the corresponding side-band. Nevertheless, the final steady-state is determined by the simultaneous action of all the sub-baths. Importantly, the contribution of a particular sub-bath depends on its bath spectral response function $G(\omega)$ and a weight factor $\Pq$. While the former encodes all the information regarding the physical baths and is therefore often pre-determined and difficult to tune, the later depends on the characteristics of the modulation itself. In this work, we illustrate that in contrast to a symmetric modulation, an asymmetric pulse can provide a more flexible control on the final steady state, either through a `renormalization' of the excitation energies that are dynamically generated in the steady state, or by lifting the symmetry of the weight factors $\Pq$ about $q=0$. 

Let us briefly recapitulate some of the well-known properties of continuous quantum thermal machines. A continuous quantum thermal machine, which usually consists of  a two (or few)-level system (TLS) perpetually coupled to two thermal baths, is known to be capable of working both as a quantum heat engine and quantum refrigerator \cite{kosloff13, kurizki15, kurizki13}. The mode of operation is determined and controlled by the modulation frequency; at the critical value of the modulation frequency where the system switches the mode of operation, the system attains the Carnot efficiency \cite{carnot}, although all heat currents as well as the power generated vanish. This is similar in characteristic to the classical Otto engine, which also achieves the Carnot efficiency in the limit of vanishing work output. We note that apart from engine-like and refrigerator-like operations, quantum thermal machines are also known to operate as `heaters' where an external work is used to supply heat energy to both the hot and cold baths, and as `accelerators' where the heat transfer from the hot bath to the cold bath is accelerated with the help of an external power source \cite{fazio15, fazio16, buffoni19}.

Apart from functioning as toy models of quantum heat engines and refrigerators, simple quantum systems, such as a qubit (TLS) or a harmonic oscillator, when connected to baths, have also found applications as efficient `quantum probes' in the field of quantum metrology \cite{maccone11, degen17, zwick17, pezze18, sciarrino20}. The idea is to make indirect measurements on the system which in some cases can be more precise than a direct measurement of the small parameter to be estimated. As for example, it has been shown that performing indirect measurements at the Carnot point, where a thermal machine switches opertion from engine-like to refrigerator-like operations or vice-versa, can significantly enhance precision in quantum thermometric and magnetometric measurements \cite{brunner17, dutta20}. Recently, it has been shown that a measurement on the steady state populations of a periodically modulated quantum system in conjunction with a thermal bath of unknown temperature, provides a more precise estimation of the bath temperature \cite{mukherjee19}. Importantly, the quantum Fisher information (QFI) using this protocol scales as $1/T^2$; consequently, the theoretical lower bound of the relative error set by the quantum Cramer-Rao bound is rendered independent of temperature, which allows measurements with finite error up to extremely low temperatures.

The rest of the paper is organized as follows. In Sec.~\ref{sec_asym}, we briefly outline the Floquet-Lindblad formalism  used for analyzing dynamical evolution of open quantum systems modulated periodically with an APM and also derive the general form of the steady state. In Sec.~\ref{sec_eng}, we explore the operation of a continuous thermal machine constructed by modulating a two-level system with an APM and perpetually coupled with two thermal baths. In Sec.~\ref{sec_thermo}, we highlight the advantage of using an APM in enhancing precision in thermometric measurement using dynamical control. Concluding statements and scope for future research are presented in Sec.~\ref{sec_con}.   A short derivation of the heat currents found in the steady state of the continuous quantum thermal machine is outlined in Appendix~\ref{app_curr}.

\section{Asymmetric-pulse modulation of open quantum systems}\label{sec_asym}

\begin{figure}
	\includegraphics[width=\columnwidth]{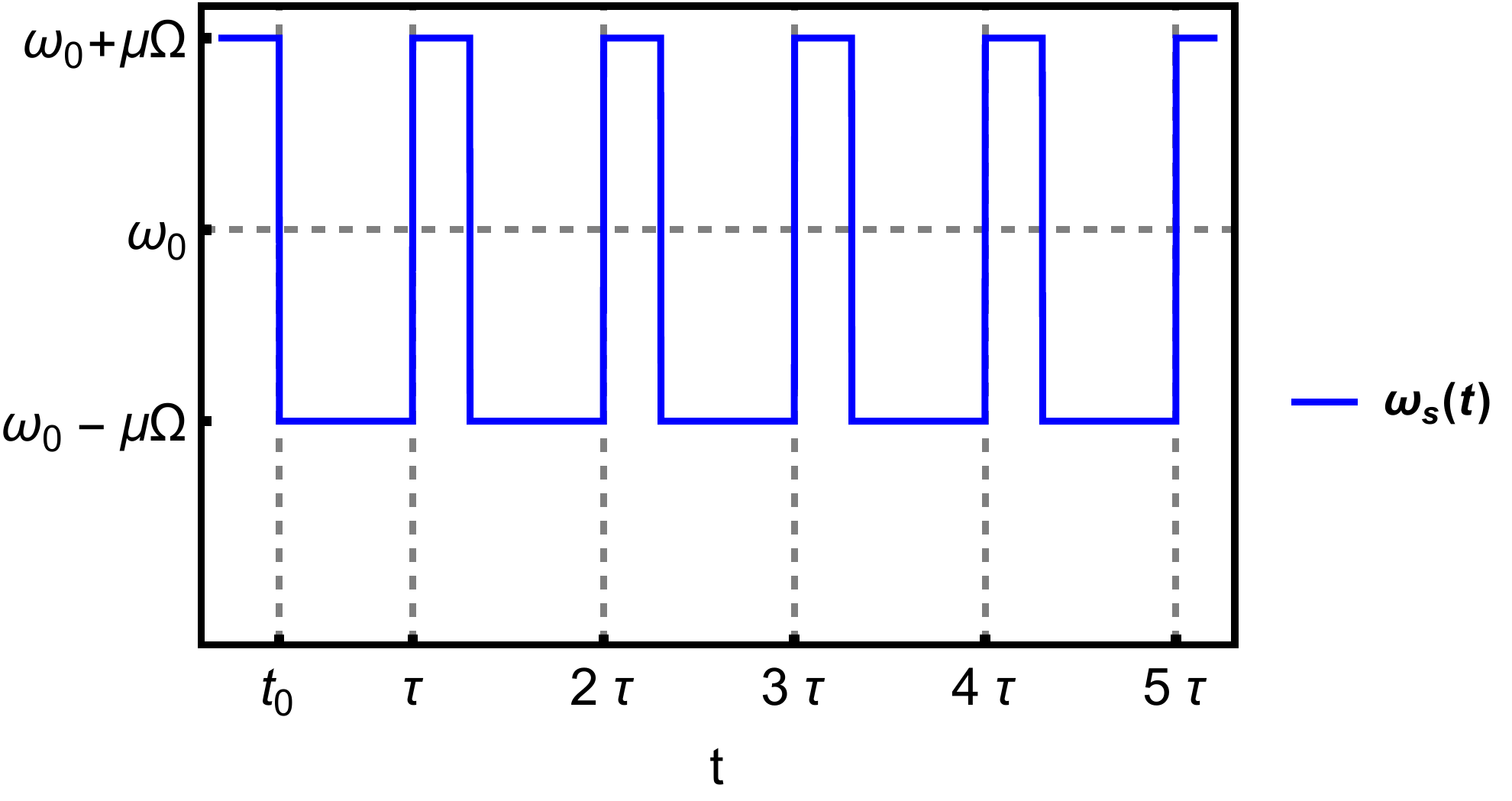}
	\caption{Profile of the asymmetric pulse modulation (APM), with time-period $\tau=2\pi/\Omega$ and modulation strength $\mu$, applied on the Hamiltonian of the TLS (see Eq.~\eqref{eq_hs}). Within a single time-period, the pulse is `up' for a duration $t_0$ and `down' for  a duration $\tau-t_0$. For $t_0=\tau/2$, the pulse is symmetric. }\label{fig_asym_pulse}
\end{figure}

Let us consider a two-level system (TLS) coupled to an arbitrary number of  thermal baths which do not interact with each other. The Hamiltonian of the composite system (including the baths) is,
\begin{equation}
H(t) = H_s(t)+\sum_b H_b+H_I,
\end{equation}
where $H_s(t)$ is the periodically modulated Hamiltonian of the system, $H_b$ is the Hamiltonian describing the $b$-{th} thermal bath and $H_I$ denotes the time-independent interaction between the system and the baths. The system Hamiltonian $H_s(t)$ is of the form (see Fig.~\ref{fig_asym_pulse}),
\begin{align}\label{eq_hs}
H_s(t) =\frac{1}{2}\omega_s(t)\sigma_z=& \frac{1}{2}\left(\omega_0+\mu\Omega\right)\sigma_z, \hspace{0.75cm}0<t<t_0,\non\\
	   & \frac{1}{2}\left(\omega_0-\mu\Omega\right)\sigma_z, \hspace{0.75cm}t_0<t<\tau,
\end{align}
where $\Omega=2\pi/\tau$ is the frequency of modulation, $\mu$ determines the modulation strength and $\sigma_z$ is a Pauli matrix. We note that the above modulation corresponds to a symmetric pulse when $t_0=\tau/2$; for $t_0\neq \tau/2$, it corresponds to an APM. We do not specify any explicit form of $H_b$ as we are only interested in the reduced dynamics of the system. As for the interaction $H_I$, the following form is assumed,

\begin{equation}\label{eq_int}
H_I=\lambda\sum_{b} \sigma_x\otimes B_b,
\end{equation} 
where $\lambda$ denotes the strength of system-bath coupling, which is assumed to be identical in case of all baths while  $\sigma_x$  (Pauli matrix) and $B_b$ are Hermitian operators acting locally on the Hilbert spaces of the system and $b$-{\rm th} bath, respectively.

To analyze the dynamical evolution, we first note that the Floquet evolution operator which drives the evolution over one period in the absence of dissipation is given by,
\begin{subequations}
\begin{equation}
U_F=\mathcal{T}e^{-i\int_0^\tau H_s(t)dt}=e^{-iH_F\tau},
\end{equation}
where the Floquet Hamiltonian $H_F$ is given by,
\begin{equation}\label{eq_flo_ham}
H_F=\frac{1}{2}\left(\omega_0+\mu\Omega\left(\frac{2t_0}{\tau}-1\right)\right)\sigma_z=\frac{1}{2}\bar{\omega}\sigma_z.
\end{equation}
\end{subequations}
We note here that for symmetric modulation $t_0=\tau/2$, the Floquet Hamiltonian is simply $H_F=\omega_0\sigma_z/2$, i.e. the Floquet spectrum is insensitive to the strength $\mu$ or the frequency $\Omega$ of modulation. Using the Floquet Hamiltonian $H_F$, the time-evolution operator for the system can be recast as,
\begin{equation}
U_s(t)=U_s(t)e^{iH_Ft}e^{-iH_Ft}=R(t)e^{-iH_Ft},
\end{equation} 
where one can verify that $R(t+\tau)=R(t)$. The periodicity of the operator $R(t)$ permits a Fourier expansion of $U_s(t)$ as follows,
\begin{subequations}
\begin{equation}
U_s(t)=\sum_qR_qe^{-iq\Omega t}e^{-iH_Ft},
\end{equation}
\begin{equation}\label{eq_rq}
R_q=\frac{1}{\tau}\int_0^\tau R(t)e^{iq\Omega t}dt.
\end{equation}
\end{subequations}
We can now evaluate Eq.~\eqref{eq_int} in the interaction picture as,
	\begin{align}\label{app_eq_int}
	\tilde{H}_I &=\lambda\sum_{b=1}^k \left(U_s^\dagger(t)\sigma_xU_s(t)\right)\otimes \left(U_b^\dagger(t)B_bU_b(t)\right)\non\\
	&=\lambda\sum_{b=1}^k\sigma_x(t)\otimes B_b(t)
	\end{align}
	where, 
\begin{multline}
\sigma_x(t)=\sum_{q,q'}\Big[R_{q'}^\dagger\sigma^{-}R_qe^{-i\bar{\omega}t-i(q-q')\Omega t}\\+R_{q'}^\dagger\sigma^{+}R_qe^{i\bar{\omega}t-i(q-q')\Omega t}\Big].
\end{multline}
Substituting $R_q$ from Eq.~\eqref{eq_rq} in the above equation, we obtain,
\begin{equation}\label{eq_sigma_xt}
\sigma_x(t)=\sum_{q\in\mathbb{Z}}\left(\xi_qe^{-i(\bar{\omega}+q\Omega)t}\sigma^-+\xi_q^*e^{i(\bar{\omega}+q\Omega)t}\sigma^+\right),
\end{equation}
where,
\begin{multline}\label{eq_xiq}
\xi_q=\frac{1}{\tau}\Big(\int_0^{t_0}e^{-i\int_0^t(\omega_0+\mu\Omega-\bar{\omega})dt'}e^{-iq\Omega t}dt\\+\int_{t_0}^{\tau}e^{\left[-i\int_0^{t_0}(\omega_0+\mu\Omega-\bar{\omega})dt'+\int_{t_0}^t(\omega_0-\mu\Omega-\bar{\omega})dt'\right]}e^{-iq\Omega t}dt\Big)\\
=\frac{\mu \left(e^{-2i\mu\Omega t_0(1-\frac{t_0}{\tau})+iq\Omega t_0}-1\right)}{i\pi\left(q-2\mu(1-\frac{t_0}{\tau})\right)\left(q+2\mu\frac{t_0}{\tau}\right)}.
\end{multline}

We are now in a position to derive the dynamical equation of motion for the system following Refs.~[\onlinecite{kosloff13, kurizki15, kurizki13}]. We begin with the von-Neumann equation of motion in the interaction picture,
\begin{equation}
\frac{d\tilde{\rho}_{tot}}{dt}=-i\left[\tilde{H}_I,\tilde{\rho}_{tot}\right]
\end{equation} 
where $\tilde{\rho}_{tot}$ is the density matrix of the composite system. Under the weak-coupling coupling approximation ($\lambda\ll 1$), the system and the baths are assumed to exist in a direct product state at all times and the system has no significant back-action on the baths, i.e. $\tilde{\rho}_{tot}(t)=\tilde{\rho}(t)\otimes\rho_{B,1}\otimes\rho_{B.2}\otimes\dots$, where $\rho_{B,b}$ denotes the stationary density matrix of the $b^{th}$ bath. This allows us to extract the dynamics of $\tilde{\rho}(t)$ as,
\begin{multline}
\frac{d\tilde{\rho}(t)}{dt}=-i\lambda\sum_b\Tr_b\left[\sigma_x(t)\otimes B_b(t),\tilde{\rho}(0)\otimes\rho_{B,b}\right]\\-\lambda^2\sum_b\int_0^t\Tr_b\left[\sigma_x(t)\otimes B_b(t),\left[\sigma_x(t')\otimes B_b(t'),\tilde{\rho}(t')\otimes\rho_{B,b}\right]\right]dt',
\end{multline}
where $\Tr_b[.]$ denotes a partial trace over the degrees of freedom of the $b^{th}$ bath. The first term in the r.h.s of the above equation vanishes if $\la B_b(t)\ra=\Tr(\rho_{B,b}B(t))=0$, which is often the case and is also assumed to be true in the rest of our work. Next, we assume that the dynamics is Markovian or memory-less in nature. The  Markovian nature is ensured when the time scale of relaxation of the system is much higher than the decay time of the bath correlations $\tau_b$. As a result, the evolution at a given time $t$ is independent of the time preceding $t$. On substituting $t'=t-s$ in the above equation, we can therefore let the integral run from $s=0$ to $s=\infty$ as the integrand vanishes under the Markovian approximation for $s\gg\tau_b$, i.e.,
\begin{multline}\label{eq_lind_2}
\frac{d\tilde{\rho}(t)}{dt}=-\lambda^2\sum_b\int_0^\infty\Tr_b\Big[\sigma_x(t)\otimes B_b(t),\\\left[\sigma_x(t-s)\otimes B_b(t-s),\tilde{\rho}(t)\otimes\rho_{B,b}\right]\Big]ds
\end{multline}
where we have also assumed $\tilde{\rho}(t-s)=\tilde{\rho}(t)$ under the Markovian approximation. Substituting $\sigma_x(t)$ from Eq.~\eqref{eq_sigma_xt} and neglecting fast oscillating terms (secular approximation) of the form  $e^{\pm2i\bar{\omega}t}$, $e^{iq\Omega t}$ and $e^{i(\pm 2\bar{\omega}+q\Omega)t}$ where $q\neq0$, we arrive at,
\begin{multline}
\frac{d\tilde{\rho}(t)}{dt}=\sum_{b,q}P_q\Big[\Gamma^{b}(\bar{\omega}+q\Omega)
\Big(\sigma^-\tilde{\rho}(t)\sigma^+-\sigma^+\sigma^-\tilde{\rho}(t)\Big)\\
+\Gamma^{b}(-\bar{\omega}-q\Omega)
\Big(\sigma^+\tilde{\rho}(t)\sigma^--\sigma^-\sigma^+\tilde{\rho}(t)\Big]+h.c.
\end{multline}
where,
\begin{equation}\label{eq_Pq}
P_q=|\xi_q|^2=\frac{4\mu^2\sin^2\left[2\mu\pi\frac{t_0}{\tau}\left(1-\frac{t_0}{\tau}\right)-q\pi\frac{t_0}{\tau}\right]}{\pi^2\left[q-2\mu\left(1-\frac{t_0}{\tau}\right)\right]^2\left[q+2\mu\frac{t_0}{\tau}\right]^2}
\end{equation} 
and
\begin{align}
\Gamma^b(\bar{\omega}+q\Omega)&=\lambda^2\int_0^\infty ds e^{i(\bar{\omega} +q\Omega)s}\la B_b^\dagger(t)B_b(t-s)\ra_b\nonumber\\
&=\lambda^2\int_0^\infty ds e^{i(\bar{\omega}+q\Omega) s}\la B_b^\dagger(s)B_b(0)\ra_b.
\end{align}

Finally, separating the real and the imaginary parts of the function $\Gamma^b(\bar{\omega}+q\Omega)$ as
\begin{equation}
\Gamma^b(\bar{\omega}+q\Omega)=\frac{1}{2}G^b(\bar{\omega}+q\Omega)+iS^b(\bar{\omega}+q\Omega),
\end{equation} 
where $G^b$ is usually known as the bath spectral function of $b$-th bath, we obtain the Floquet-Lindblad equation (in the interaction picture) governing the evolution of the reduced density matrix of the TLS as,
\begin{equation}\label{eq_lindblad}
\frac{d\tilde{\rho}}{dt}=\mathcal{L}\tilde{\rho}=-\sum_{b,q}i\left[H^{b,q}_{LS},\tilde{\rho}\right]+\sum_{b,q}\mathcal{L}^b_q\tilde{\rho},
\end{equation}
where 
\begin{subequations}
\begin{equation}
H^{b,q}_{LS}=P_q\left(S^b(\bar{\omega}+q\Omega)\sigma^+\sigma^-+S^b(-\bar{\omega}-q\Omega)\sigma^-\sigma^+\right),
\end{equation}
\begin{multline}
\mathcal{L}^b_q=P_q\Big[G^b(\bar{\omega}+q\Omega)\left(\sigma^-\rho\sigma^+-\frac{1}{2}\{\sigma^+\sigma^-,\rho\}\right)\\+G^b(-\bar{\omega}-q\Omega)\left(\sigma^+\rho\sigma^--\frac{1}{2}\{\sigma^-\sigma^+,\rho\}\right)\Big].
\end{multline}
\end{subequations}
The Hamiltonian $H^{b,q}_{LS}$ renormalizes the system Hamiltonian in the Schrodinger picture, inducing the so-called Lamb corrections \ct{breuer02}. In the rest of this work, we ignore these corrections as they are small in magnitude and does not affect the dynamics of the steady-state.

The above equation can be interpreted as follows. The periodic modulation results in the generation of multiple side-bands or photon sectors of the system Hamiltonian with energy-gap $\bar{\omega}$ and are separated by integer multiples $q$ of the modulation frequency $\Omega$. {The interaction with the bath induces excitations between the side-bands} and the resulting dissipation is captured by the Lindblad operators $\mathcal{L}_q^b$. {Importantly, we note that  the energy-gap $\bar{\omega}$ depends both on $\mu$ and the fraction of up time $t_0/\tau$ (see Eq.~\eqref{eq_flo_ham}) in the case of APM, unlike that in the case of the symmetric pulse where $\bar{\omega}=\omega_0$. Secondly, the weight of the different side-bands $P_q$ is no longer symmetric about $q=0$ and can be tuned by changing the fraction of up (or down) time $t_0/\tau$. We will shortly illustrate the potential advantages of these results in the next couple of sections.} Lastly, we note that the asymptotic steady state is now easily obtained by solving the eigen equation $\mathcal{L}\tilde{\rho}_{ss}=0$ as,
\begin{subequations}\label{eq_steady}
\begin{equation}
\tilde{\rho}_{ss}=\frac{1}{1+r}\begin{pmatrix}
			r & 0\\0 & 1\end{pmatrix},
\end{equation}

where,
\begin{equation}\label{eq_r_ori}
r=\frac{\sum_{q,b}\Pq G^b\left(\bar{\omega}+q\Omega\right)e^{-\frac{\bar{\omega}+q\Omega}{T_b}}}{\sum_{q,b}\Pq G^b\left(\bar{\omega}+q\Omega\right)},
\end{equation} 
\end{subequations}
and $T_b$ is the temperature of the $b$-{th} bath.  It can be shown that the time-independent steady state $\tilde{\rho}_{ss}$ in the interaction picture translates to a periodic steady state $\rho_{ss}$ in the Schrodinger picture which satisfies $\rho_{ss}(t+\tau)=\rho_{ss}(t)$. However, in the rest of this work, we will analyze the relevant quantities only in the interaction picture.

\begin{figure*}
	\centering
	\subfigure[]{
		\includegraphics[width=0.45\textwidth]{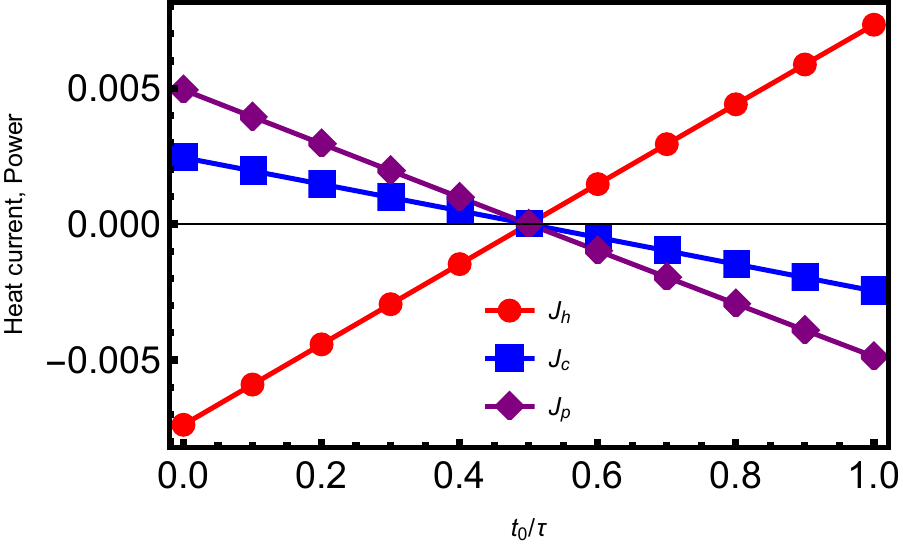}
		\label{fig_ref_eng}}	
	\centering
	\subfigure[]{
		\includegraphics[width=0.45\textwidth]{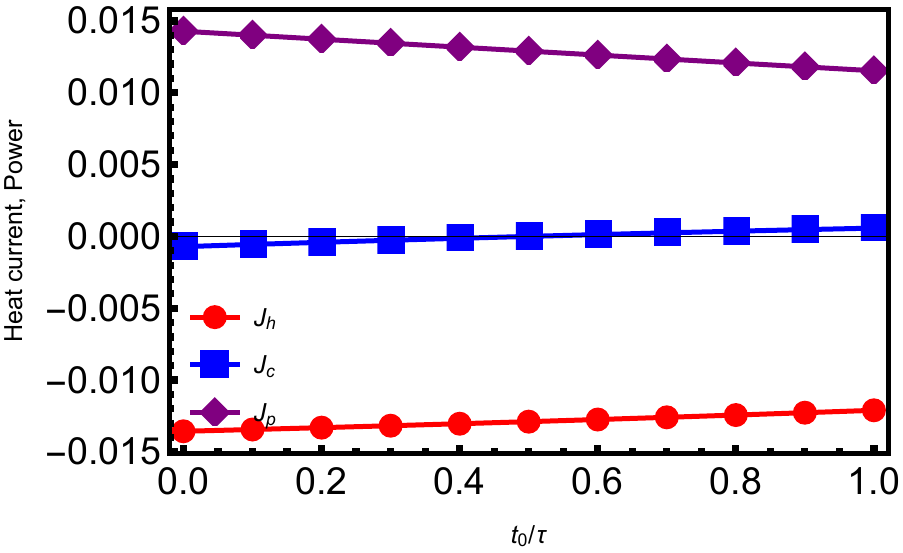}
		\label{fig_ref_heat}}
	\caption{(a) Illustration of the refrigerator to engine transition with parameters chosen as $\omega_0=2$, $T_c=1$, $T_h=3$, $\Omega=1$ and $\mu=0.01$. The Carnot point, as can be calculated from Eq.~\eqref{eq_carnot_asym}, is therefore $t_0/\tau=0.5$. At the Carnot point,   the heat currents and power vanish simultaneously and the transition is evident from the flipping of signs of all the currents across the point. (b) An example of heater to refrigerator transition  for parameters chosen as $\Omega=\omega_0=0.1$, $\mu=0.1$ and same temperature of the baths as in (a). Unlike (a), this transition always occurs at   $t_0/\tau=0.5$; across this point, only of sign of  the cold current $J_c$ changes thereby signaling a refrigerator to heater transition.}
\end{figure*}

\section{APM in continuous quantum thermal machines}\label{sec_eng}

{We now consider the working of a simple continuous thermal machine  which consists of a TLS  coupled to two thermal baths}. The `hot' bath has a temperature $T_h$ and the cold one has temperature $T_c$, such that $T_h>T_c$. {Further, the TLS is periodically modulated with an APM of the form given in Eq.~\eqref{eq_hs}.} In steady state operation, energy currents flow continuously between the TLS and the hot and cold baths, which are identified as the hot current $J_h$ and cold current $J_c$, respectively. In addition, the continuous pumping of energy in or out of the system through the modulation is identified as the work current or power $J_p$. In analogy with classical machines, the thermal machine is considered to operate as a heat engine when $J_h>0$, $J_c<0$ and $J_p<0$. In the refrigeration regime of operation, the quantities reverse their sign. In the steady state defined in Eq.~\eqref{eq_steady}, the heat currents are calculated as (see Appendix~\ref{app_curr}),

\begin{equation}\label{eq_heat}
J_{h(c)}=\sum_q\frac{\bar{\omega}+q\Omega}{r+1}\Pq G^{h(c)}\left(\bar{\omega}+q\Omega\right)\left(e^{-\frac{\bar{\omega}+q\Omega}{T_{h(c)}}}-r\right).
\end{equation}
The power is then calculated using the energy conservation principle as,
\begin{equation}\label{eq_power}
J_p=-(J_h+J_c).
\end{equation}

Let us  now take a closer look at the coefficients $\Pq$. {Throughout this section, we work within the  limit of weak modulation, i.e., $\mu\ll 1$. The symmetric distribution of $P_q$ around $q=0$ is restored in this limit and  Eq.~\eqref{eq_Pq} reduces to},
\begin{subequations}\label{eq_pq_thermal}
	\begin{equation}
	P_0=\left[\frac{\sin\left(\mu\Omega t_0(1-\frac{t_0}{\tau})\right)}{2\mu\pi\frac{t_0}{\tau}(1-\frac{t_0}{\tau})}\right]^2,
	\end{equation}
	\begin{equation}
	P_{q\neq 0}=\left[\frac{2\mu\sin\left(\frac{q\Omega t_0}{2}\right)}{\pi q^2}\right]^2
	\end{equation}
\end{subequations} 
The above equation shows that the value of $\Pq$ diminishes as $|q|^{-4}$ with increasing $|q|$. Therefore, we keep ourselves restricted to only the leading order coefficients $P_0$ and $P_{\pm1}$. Secondly, a `spectral separation' of the dominant modes is crucial for functioning of the thermal machine. For this purpose, we also introduce the following cutoffs for the bath spectral functions,
\begin{equation}\label{eq_spec_sep}
G^h(\omega)=0\quad \forall~\omega \leq \bar{\omega},\quad G^c(\omega)=0\quad \forall~\omega\geq \bar{\omega}
\end{equation}

Using Eqs.~\eqref{eq_heat} and ~\eqref{eq_power}, the heat currents and the power are now found to be,
\begin{widetext}
\begin{subequations}\label{eq_all_curr}
\begin{equation}
J_h=\mathcal{K}\left(\bar{\omega}+\Omega\right)\left(e^{-\frac{\bar{\omega}+\Omega}{T_h}}-e^{-\frac{\bar{\omega}-\Omega}{T_c}}\right),
\end{equation}
\begin{equation}
J_c=\mathcal{K}\left(\bar{\omega}-\Omega\right)\left(e^{-\frac{\bar{\omega}-\Omega}{T_c}}-e^{-\frac{\bar{\omega}+\Omega}{T_h}}\right),
\end{equation}
\begin{equation}
J_p=-2\mathcal{K}\Omega\left(e^{-\frac{\bar{\omega}+\Omega}{T_h}}-e^{-\frac{\bar{\omega}-\Omega}{T_c}}\right),
\end{equation}
where $\mathcal{K}$ is a positive constant given by, 

\begin{equation}
\mathcal{K}=\frac{4\mu^2\sin^2\left(\frac{\Omega t_0}{2}\right)G^h(\bar{\omega}+\Omega)G^c(\bar{\omega}-\Omega)}{\pi^2\left[G^h(\bar{\omega}+\Omega)\left(1+e^{-\frac{\bar{\omega}+\Omega}{T_h}}\right)+G^c(\bar{\omega}-\Omega)\left(1+e^{-\frac{\bar{\omega}-\Omega}{T_c}}\right)\right]}
\end{equation}
\end{subequations}
\end{widetext}
We now identify the different modes of operation as follows. {As already mentioned}, the thermal machine works as a heat engine if $J_h>0$, $J_c<0$ and $J_p<0$. Similarly, refrigeration occurs when $J_h<0$, $J_c>0$ and $J_p>0$. Additionally, the thermal machine is also capable of working as a `heater'  when $J_h<0$, $J_c<0$ and $J_p>0$. Let us first assume that $\bar{\omega}>\Omega$. Examining Eq.~\eqref{eq_all_curr}, it is clear that engine like operation is achieved when
$(\bar{\omega}-\Omega)/T_c>(\bar{\omega}+\Omega)/(T_h)$ while refrigerator like operation is achieved for $(\bar{\omega}-\Omega)/T_c<(\bar{\omega}+\Omega)/T_h$. Therefore, there exists a critical point, namely the Carnot point, where the heat currents as well as the power vanish and the thermal machine switches operation from engine like to refrigerator like and vice-versa. The Carnot point is identified by the relation,
\begin{equation}\label{eq_carnot}
\left(\frac{\Omega}{\bar{\omega}}\right)_{cr}=\frac{T_h-T_c}{T_h+T_c}.
\end{equation}

Note that for a symmetric pulse $t_0=\tau/2$, the above equation reduces to $\Omega_{cr,sym}=\omega_0(T_h-T_c)/(T_h+T_c)$, which is identical for the case of sinusoidal modulation in Ref.~[\onlinecite{kurizki13}]. In this case, the mode of operation of the thermal machine can only be switched by tuning the modulation frequency $\Omega$, which provides the only degree of control over the mode of operation. On the contrary, in the case of the APM, one can rearrange Eq.~\eqref{eq_carnot} as,
\begin{equation}\label{eq_carnot_asym}
\left(\frac{t_0}{\tau}\right)_{cr}=\frac{1}{2}\left[1+\frac{\omega_0}{\mu}\left(\frac{1}{\Omega_{cr,sym}}-\frac{1}{\Omega}\right)\right],
\end{equation}
which implies that for a fixed modulation frequency $\Omega$, the transition can also be induced by tuning $t_0$, as shown in Fig.~\ref{fig_ref_eng}. In other words, the the ratio of up time duration to the total pulse duration, i.e., $t_0/\tau$, provides an extra degree of control over the mode of operation of the thermal machine. However, we note that since $0< t_0/\tau< 1$, a critical $(t_0/\tau)_{cr}$ exists only if the following condition is satisfied,
\begin{equation}
\left|\frac{\omega_0}{\mu}\left(\frac{1}{\Omega_{cr,sym}}-\frac{1}{\Omega}\right)\right|\leq 1.
\end{equation}

Next, let us further consider the situation $\Omega=\omega_0$. We therefore have,
\begin{subequations}
	\begin{equation}
 \bar{\omega}+\Omega=2\omega_0\left(1+\mu\left(\frac{t_0}{\tau}-\frac{1}{2}\right)\right),
\end{equation}
\begin{equation}
\bar{\omega}-\Omega=2\mu\omega_0\left(\frac{t_0}{\tau}-\frac{1}{2}\right).
\end{equation}
\end{subequations}
If we now choose $t_0/\tau<1/2$, we have $\bar{\omega}+\Omega>0$ as $\mu\ll 1$, while $\bar{\omega}-\Omega<0$. Consequently, $e^{-(\bar{\omega}-\Omega)/T_c}>e^{-(\bar{\omega}+\Omega)/T_h}$. With this choice of parameters, it can be easily seen from Eq.~\eqref{eq_all_curr} that the heat currents satisfy $J_h, J_c <0$ and the power $J_p>0$. As already mentioned, this corresponds to the thermal machine working as a heater where work is done to supply heat to both the baths. On the other hand, for $t_0/\tau>1/2$, we have $\bar{\omega}-\Omega>0$ but, the 
condition $e^{-(\bar{\omega}-\Omega)/T_c}>e^{-(\bar{\omega}+\Omega)/T_h}$ still holds as $\bar{\omega}+\Omega\gg\bar{\omega}-\Omega\approx 0$. Hence, only the sign of $J_c$ is reversed, which corresponds to a refrigerator like operation. An example of this transition is illustrated in Fig.~\ref{fig_ref_heat}. Thus, we have demonstrated the possibility of tuning $t_0$ to drive a refrigerator-heater transition if the modulation frequency is in resonance with the un-modulated TLS, i.e. $\Omega=\omega_0$.

Before concluding this section, we note that a weak modulation $\mu\ll1$ is assumed in our analysis to ensure that the coefficients $P_q$ remain symmetric about the center band $q=0$ (see Eq.~\eqref{eq_pq_thermal}). This simplifies the analysis and allows us to focus on the other consequence of the APM, i.e. the renormalization of the energy-gap of the side bands. This renormalization, as we have seen, provides an extra control over the mode of operation of the thermal machine. We further note that a shift in the distribution of $P_q$, resulting from large modulation strengths,  may lead to situations in which the dominant modes do not remain spectrally separated, which in turn may lead to unstable machine operation. This is because the spectral separation, as we recall, are engineered by cut-offs in the bath spectra (see Eq.~\eqref{eq_spec_sep} for example) and are therefore difficult to tune for all practical purposes. Nevertheless, if the spectral separation is appropriately adjusted, we expect our results to remain valid for moderate to high strengths of $\mu$ as well, since the renormalization of the energy gaps occur for all strengths of modulation.

\section{APM in quantum thermometry}\label{sec_thermo}

We now illustrate how a TLS modulated with an APM can be used to measure the temperature of a thermal bath with a precision higher than that possible with a symmetric pulse. The advantage of using the APM is manifested in the form of a higher QFI and consequently a lower minimum bound on relative error. We briefly outline the measurement protocol below.

\begin{figure*}
	\centering
	\subfigure[]{
		\includegraphics[width=0.45\textwidth]{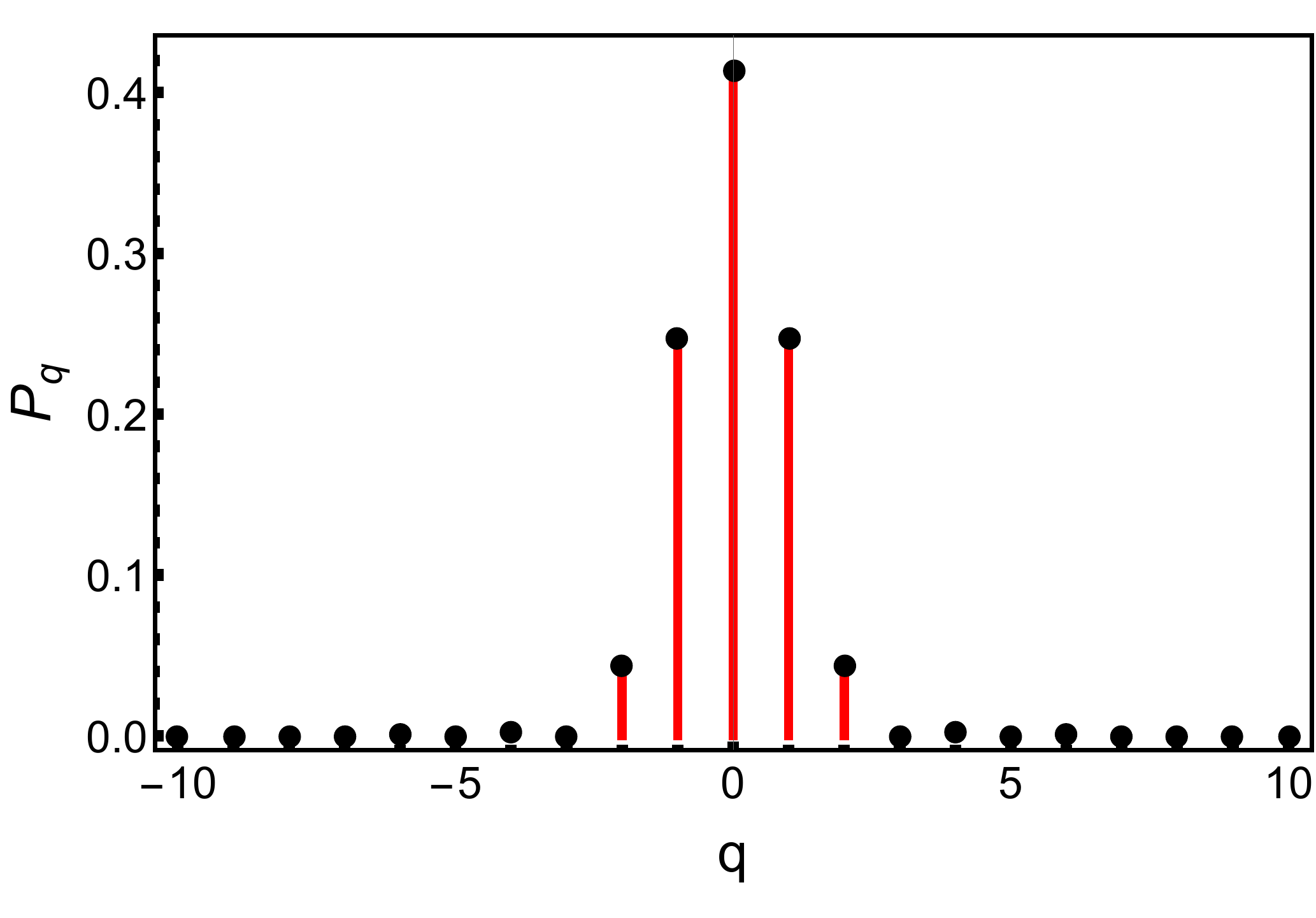}
		\label{fig_pq_sym}}	
	\centering
	\subfigure[]{
		\includegraphics[width=0.45\textwidth]{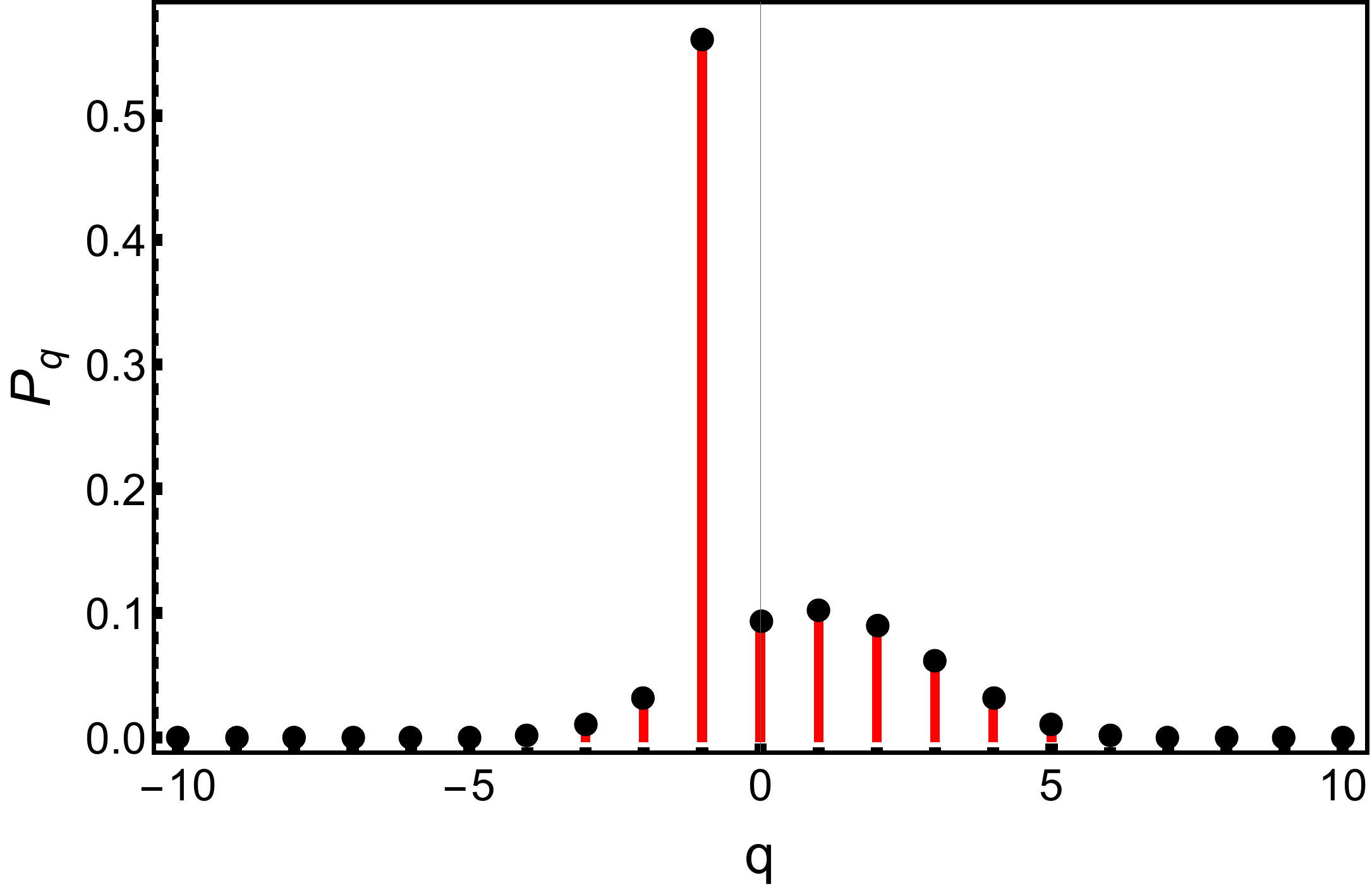}
		\label{fig_pq_asym}}
	\caption{Weight of the Floquet side-bands for (a) $t_0/\tau=0.5$, $\mu=1$ and (b) $t_0/\tau=0.25$, $\mu=2$. One can clearly see that in the later asymmetric case $P_{-1}\gg P_{n\neq-1}$.}\label{fig_Pq}.
\end{figure*}

Let us consider a thermal bath whose temperature $T$ is to be measured. The quantum probe we choose is a TLS which is coupled to the bath and its energy levels are periodically modulated using an APM as discussed in Sec.~\ref{sec_asym}, with the number of baths now restricted to one. The TLS therefore reaches a thermal steady state $\tilde{\rho}_{ss}$  (see Eq.~\eqref{eq_steady}). This steady state is characterized through the parameter $r$ given by, 
\begin{equation}\label{eq_r_thermo}
r=\frac{\sum_{q}\Pq G\left(\omega_q\right)e^{-\frac{\omega_q}{T}}}{\sum_{q}\Pq G\left(\omega_q\right)}.
\end{equation}
where $\omega_q=\bar{\omega}+q\Omega$. First, we note that for given values of $\mu$ and $t_0/\tau$, we can always set the modulation frequency $\Omega$, such that $\omega_{-n}=\bar{\omega}-n\Omega\to0^+$ where $n>0$ and $\omega_{-n}\ll\omega_{-(n-1)}, ~\omega_{-(n-2)},\dots$ . Further, if  $T\sim\omega_{-n}$ , we have $e^{-\omega_{-(n-1)}/T}, e^{-\omega_{-(n-2)}/T},\dots\ll e^{-\omega_{-n}/T}$. We can therefore neglect contributions from all side-bands with $\omega_{q>-n}$ in the numerator. In addition, we choose $\omega_{-n}$ such that $P_{q<-n}\ll P_{-n}$, i.e., we assume that the coefficients $P_q$ of all the side-bands with $\omega_q<0$ are suppressed. Hence, we can also neglect contributions from all side-bands with $\omega_{q<-n}$ in both the numerator as well as the denominator. Consequently, Eq.~\eqref{eq_r_thermo} simplifies to,
\begin{equation}\label{eq_rn}
r\approx\frac{P_{-n} G\left(\omega_{-n}\right)e^{-\frac{\omega_{-n}}{T}}}{\sum_{q=-n}^\infty\Pq G\left(\omega_q\right)}=\nu e^{-\frac{\omega_{-n}}{T}},
\end{equation}
where $P_q$ is obtained from Eq.~\eqref{eq_Pq}. The steady state therefore assumes the form (see Eq.~\eqref{eq_steady}).
\begin{equation}
\tilde{\rho}_{ss}=\begin{pmatrix}
\varrho_{1} & 0\\
0 & \varrho_{2}
\end{pmatrix}		
= \frac{1}{1+\nu e^{-\frac{\omega_{-n}}{T}}}\begin{pmatrix}
\nu e^{-\frac{\omega_{-n}}{T}} & 0\\
0 & 1
\end{pmatrix}.
\end{equation}
One now measures the steady state populations $\varrho_{1(2)}$, which depend on $\omega_{-n}$ and $T$. The former is calculated using the relation $\omega_{-n}=\bar{\omega}-n\Omega$, where $\bar{\omega}$ and $\Omega$ experimentally controlled quantities. Hence, one can infer the temperature $T$ by measuring the steady state populations.
	
However, it is known that the minimum error in any such indirect measurement is theoretically lower bounded by the quantum Cramer-Rao bound which states,
\begin{equation}\label{eq_bound}
\frac{\Delta T}{T}\geq\varepsilon=\frac{1}{T\sqrt{\mathcal{M}\mathcal{H}}},
\end{equation}
where $\mathcal{M}$ is the number of repeated measurements performed and $\mathcal{H}$ is the quantum Fisher information (QFI) \cite{caves94, paris09}. In our case, the relevant QFI is easily calculated as,
\begin{equation}
\mathcal{H}=\sum_{i=1}^2\frac{1}{\varrho_i}\left|\frac{\partial\varrho_i}{\partial T}\right|^2=\frac{\nu e^{-\frac{\omega_{-n}}{T}}\omega_{-n}^2}{\left(1+\nu e^{-\frac{\omega_{-n}}{T}}\right)^2T^4},
\end{equation}
where $T\sim\omega_{-n}$. 

It is easy to check that the maximum QFI is obtained for $\nu=e^{\omega_{-n}/T}$. At $T\approx\omega_{-n}$, this corresponds to $\nu=2.72$. However, by definition, $\nu\leq1$ and hence optimality is achieved for $\nu=1$ when the following condition is satisfied,
\begin{equation}\label{eq_cond_therm}
P_{-n}G(\omega_{-n})\gg P_{-q}G(\omega_{-q}),~~q\neq n.
\end{equation}
To demonstrate how this can be easily achieved using the APM, let us consider the simple case of a nearly flat bath spectrum, 
\begin{equation}\label{eq_nfbs}
G(\omega\geq\omega_{min}>0)=G_0;~~G(\omega\to 0)=0, 
\end{equation}
Recalling Eq.~\eqref{eq_Pq}, one can check that $P_{-q}$ achieves maximum when $2\mu t_0/\tau=q$, for $q>0$. As an example, consider $\omega_{-n}=\omega_{-1}\geq\omega_{min}$. $P_{-1}$ is therefore maximum when $\mu=\tau/(2t_0)$.


To illustrate the advantage of using an APM, let us first consider the case of a symmetric pulse ($t_0=\tau/2$). In this case, the maximum value of $P_{-1}$ is therefore achieved for $\mu=1$; however $P_{-1}=0.25\leq P_0, P_1$ as is seen in Fig.~\ref{fig_pq_sym}.  This reflects the fact that for a symmetric pulse, $\Pq=\Pqm$, $\forall q$. Consequently, the only way to satisfy the condition given in Eq.~\eqref{eq_cond_therm} in the case of the symmetric pulse is through careful manipulation of the bath spectral function, for example, by setting a low upper cutoff for $G(\omega)$. We note here that the above results obtained for the symmetric pulse are similar to that obtained for a sinusoidal modulation analysed in Ref.~\onlinecite{mukherjee19}.
Let us now analyze how the scenario changes in the case of APM. For $t_0\neq \tau/2$, one can immediately see that $\Pq\neq\Pqm$. This creates the possibility of manipulating the parameters $\mu$ and $t_0$ such that $P_{-n}>P_q$ for $q\neq -n$. We illustrate this with an example where we choose  $\mu=2$ and $t_0/\tau=0.25$ which ensures that the condition $\mu=\tau/(2t_0)$ is satisfied for obtaining maximum $P_{-1}$. As shown in Fig.~\ref{fig_pq_asym}, for these choice of parameters, we have $P_{-1}\approx(1-t_0/\tau)^2=0.563\gg P_{n\neq-1}$.

To quantify the advantage, let us compare the maximum of the QFI achieved for the APM to that of the symmetric case. In the latter case, we consider only the dominant contributions arising from $P_0$ and $P_{\pm1}$, (see Fig.~\ref{fig_pq_sym}). Therefore, the parameter $\nu$ evaluates to,
\begin{equation}
\nu_{sym}=\frac{P_{-1}G(\omega_{-1})}{\sum_{q=-1}^1P_{q}G(\omega_{q})}\approx0.277,
\end{equation}
which yields a QFI of
\begin{equation}
\mathcal{H}_{sym}=\frac{0.09}{T^2},
\end{equation} 
in the limit $\frac{\omega_{-1}}{T}\approx 1$. On the other hand, for the APM, the dominant contributions arise from $~P_{-1}, ~P_0, ~P_1, ~P_2$ and $P_3$. Proceeding as before, we obtain
\begin{equation}
\nu_{asym}\approx0.621,
\end{equation} 
and the corresponding QFI value as,
\begin{equation}
\mathcal{H}_{asym}=\frac{0.18}{T^2}.
\end{equation}
Substituting the QFIs calculated above in Eq.~\eqref{eq_bound}, we finally obtain,
\begin{equation}
\frac{\varepsilon_{asym}}{\varepsilon_{sym}}\approx0.71,
\end{equation}
thereby clearly demonstrating that an APM lowers the minimum error bound as compared to the case of symmetric pulse. 

We would like to remark here that while calculating the above results, we have neglected all coefficients $P_{q<-1}$. In general, this is strictly valid only for exponential suppression of the coefficients $P_{q<-1}$, which is not the case for the symmetric or asymmetric pulse modulations. However, an exponential suppression is known to occur for other forms of modulation such as in (symmetric) sinusoidal modulations \cite{mukherjee19} and hence the desired suppression may be realized by extending our results for pulse modulation to appropriate forms of asymmetric modulations. 

Although we have demonstrated the advantage of using APM in the case of a nearly flat bath spectrum characterized by Eq.~\eqref{eq_nfbs} and $T\sim\omega_{-n}$, the results remain valid for any general spectrum as long as the condition \eqref{eq_cond_therm} is satisfied. In fact, the use of the APM becomes indispensable  for spectra in which $G(\omega_{q>-n})>G(\omega_{-n})$ for $n>0$. In such cases, the only way \eqref{eq_cond_therm} can be satisfied  is if $P_{-n}\gg P_{q\neq-n}$, which is generally not the case for symmetric modulations. However, the secular or the rotating wave approximation assumed while deriving the Floquet-Linblad equation in Sec.~\ref{sec_asym}, requires   that the thermalization time $\tau_{th} \propto G(\omega)^{-1}$ should satisfy $\tau_{th}^{-1}\ll\Omega,~\omega_{q}$. Our results are hence applicable for any arbitrary spectra as long as the secular approximation is satisfied.

\section{Conclusion}\label{sec_con}
In summary, we have analyzed the steady state dynamics of a two-level system coupled to thermal baths and periodically modulated with an asymmetric pulse. The asymmetric nature of the pulse, in addition to renormalizing the energy gap of the Floquet side-bands, also modifies the weight of the side-bands in terms of the modulation strength $\mu$ as well as the fraction of up time duration $t_0/\tau$ of the pulse. The significance of these results have been highlighted in the form of greater flexibility in the controlling the operation of continuous quantum thermal machines as well as precision enhancement in quantum thermometry, where such periodically modulated open systems have direct applications. 
	
Firstly, we have explored the consequences of using an APM on a TLS coupled to two different thermal baths. For a symmetric pulse, such a system is known to work both as a quantum heat engine and quantum refrigerator, depending upon the frequency of modulation. We have shown that tuning the duration of up time (or down time) of the asymmetric pulse, i.e., $t_0$, also allows switching of the mode of operation of the thermal machine between heat engine, refrigerator and heater regimes.  Thus, an asymmetric modulation may provide an extra degree of control over the mode of operation, which may be experimentally useful, particularly in cases when the frequency of the modulation is not easily tunable. 

Secondly, we have also shown that the control over the weights of the side-bands as provided by the parameters $\mu$ and $t_0/\tau$ finds a direct application in a quantum thermometry protocol, in which a periodically modulated TLS coupled to a thermal bath allows a precise estimation of the temperature of the bath. By tuning the weight of the Floquet side-bands with appropriate choices of $\mu$ and $t_0/\tau$, the QFI can be maximized. We have illustrated with an example that the optimality thus achieved using the APM can be superior than that possible when using a symmetric pulse modulation. This in turn leads to lowering of the theoretical minimum bound of the relative error as dictated by the quantum Cramer-Rao bound, thereby enhancing the maximum precision that can be achieved experimentally.

Finally, we would like to recall that the Floquet-Lindblad analysis presented in this work relies on the weak-coupling (Born), Markov and secular approximations. The Markov approximation, in particular, is crucial as it ensures that the TLS thermalizes to a non-equilibrium steady-state which is essential for the working of both the thermal machine and the thermometry protocol discussed in this work. Our results, as such, therefore can not be directly applied to systems undergoing non-Markovian dynamics. However, a generalization of our results to this regime may be possible after appropriate modifications in the Floquet-Lindblad framework.

As already mentioned, it might be worth exploring other forms of asymmetric modulations such as sinusoidal modulations. The generalization of our analysis to many level as well as degenerate systems, where coherence play a more significant role, may also be investigated. The consequences of using APM in the case of strong system-environment couplings would surely
turn out to be an interesting area of study.

\begin{acknowledgments}
We are grateful to A. Ghosh and V. Mukherjee for their helpful comments and suggestions. We also acknowledge S. Bandyopadhyay and S. Maity for useful discussions.  SB acknowledges CSIR, India for financial support. A.D. acknowledges financial support from SPARC program, MHRD, India.
\end{acknowledgments}

\appendix

\section{Heat currents in continuous thermal machines}\label{app_curr}
To determine the heat currents, we first note that each sub-bath in principle can individually take the system to a Gibbs-like steady state determined by the eigenvalue equation $\mathcal{L}_{q}^b\rho^{ss}_{b,q}=0$ where $b=h,c$. These steady states are of the form,
\begin{equation}
\rho_{b,q}^{ss}=\frac{1}{\mathcal{Z}}\exp{\left(\frac{\bar{\omega}+q\Omega}{\bar{\omega}}\beta_b H_{F}\right)},
\end{equation}
where $\beta_b=1/T_b$, $\mathcal{Z}=\Tr\left(\exp{\left(\frac{\bar{\omega}+q\Omega}{\bar{\omega}}\beta_b H_{F}\right)}\right)$ and $H_F$ is given by Eq.~\eqref{eq_flo_ham} of main text. Following \cite{alickitut, kurizki15}, we calculate the rate of change of von-Neumann entropy $S(t)=-\Tr\left(\rho(t)\ln\rho(t)\right)$,
\begin{equation}
\frac{d S(t)}{dt}=-\Tr\left(\dot{\rho}(t)\ln\rho(t)\right)=-\sum_{b,q}\Tr\left(\mathcal{L}_q^b\rho(t)\ln\rho(t)\right),
\end{equation}
where we have substituted Eq.~\eqref{eq_lindblad} to obtain the second equality. Next, we use Spohn's inequality, $\Tr\left(\mathcal{L}_q^b\rho(\ln\rho-\ln\rho^{ss}_{q,b})\right)\leq 0$ to arrive at,
\begin{equation}
\frac{d S(t)}{dt}\geq-\sum_{b,q}\Tr\left(\mathcal{L}_q^b\rho(t)\ln\rho^{ss}_{q,b}\right)=\sum_j\frac{J_b(t)}{T_b},
\end{equation}
where the heat currents $J_b$ in the steady state are obtained as,
\begin{equation}
J_b=\sum_q\left(\frac{\bar{\omega}+q\Omega}{\bar{\omega}}\right)\Tr\Big(\mathcal{L}_q^b\rho^{ss}H_F\Big).
\end{equation}
Finally, substituting $\rho^{ss}$ from Eqs.~\eqref{eq_steady} and~\eqref{eq_r_ori}, the heat currents assume the form,
\begin{equation}
J_{h(c)}=\sum_q\frac{\bar{\omega}+q\Omega}{r+1}\Pq G^{h(c)}\left(\bar{\omega}+q\Omega\right)\left(e^{-\frac{\bar{\omega}+q\Omega}{T_{h(c)}}}-r\right).
\end{equation}

\end{document}